\documentclass[10pt]{article}
\usepackage{graphicx}
\usepackage{latexsym,amssymb,amsmath,float}
\begin{document}

\begin{Huge}

\begin{center}

{\bf Drip Paintings and Fractal Analysis}

\end{center}

\end{Huge}

\begin{center}

Katherine Jones-Smith$^1$, Harsh Mathur$^1$ and Lawrence Krauss$^{1,2}$

\begin{small}

$^1${\em Department of Physics, Case Western Reserve University, 10900
Euclid Avenue, Cleveland OH 44106-7079, USA.}

$^2${\em Also Department of Astronomy, Case Western Reserve University.}

\end{small}

\end{center}



\vspace{2mm}

\begin{small}

{\bf 
It has been claimed \cite{abbott, taylor1,taylor2,taylor3,taylor4,taylor5}  that fractal analysis can be applied to unambiguously characterize works of art such as the drip paintings of Jackson Pollock.  This academic issue has become of more general interest following the recent discovery of a cache of disputed Pollock paintings.  We definitively demonstrate here, by analyzing paintings by Pollock and others, that fractal criteria provide no information about artistic authenticity. This work has also led to two new results in fractal analysis of more general scientific significance.  First, the composite of two fractals is not generally scale invariant and exhibits complex multifractal scaling in the small distance asymptotic limit. Second the statistics of box-counting and related staircases provide a new way to characterize geometry and distinguish fractals from Euclidean objects.}

\end{small}

\vspace{2mm}

In 1999, a highly publicized work \cite{taylor1} applied fractal analysis to the works of Jackson Pollock.  The recent discovery of a cache of approximately twenty-five paintings that may be the work of Pollock has motivated the application of these techniques in order to determine authenticity of these paintings.    However we argued in \cite{us} that in fact Pollock's drip-paintings cannot be usefully characterized as fractal, and that identical fractal characteristics can be trivially reproduced.  In this work we analyze seven drip-paintings (three by Pollock, two of the newly discovered paintings, and two commissioned works by local artists) and demonstrate conclusively that fractal criteria are not useful for authentication.  In particular, we demonstrate both that known Pollock drip-paintings, and known non-Pollock paintings meet the claimed criteria at equal levels of significance.  Our detailed analysis led us to explore fractal analysis in a more general context, motivating us to investigate the asymptotics of multifractals, and discover a new way to use fractal analysis to characterize geometry.
 
The standard technique used to determine whether an image is fractal is to cover it with a grid of square boxes of size $L$ and count the number of occupied boxes, $N$. For a fractal, $N \propto  L^{D}$   where  $D$ is the fractal dimension and is non-integer. The box-counting curve (a plot of $\log N$ against $\log L$) is therefore a straight line with slope $D$.   According to Taylor {\em et al.} \cite{taylor5} the box-counting curves of  Pollock paintings meet the following criteria: (1) there are two fractal dimensions, $ D_D < D_L$ ,where $D_D$ is the fractal dimension for boxes smaller than a transition length $L_T$ , and $D_L$ for boxes larger than $L_T$ ; (2)  $L_T  > $1.0 cm; (3) the fits to the box-counting data are low noise with $\chi  <$ 0.026; (4) for multi-colored paintings each colored layer as well as the composite painting all satisfy criteria (1-3). Fractal authentication is based on the claim that all Pollock drip paintings satisfy these criteria. In addition it is also claimed that these characterisics are exclusive to Pollock, arising from his unique mastery of chaotic motion \cite{taylor1,taylor2,taylor3,taylor4,taylor5}.

In previous work \cite{us} we identified several problems with the fractal analysis of Taylor {\em et al.}: (i) an insufficient range of box sizes was used to establish fractal behaviour; (ii) it is mathematically impossible for the visible portion of each layer  and the composite to separately behave as fractals in a multilayered painting; (iii) $\chi$ 
depends on the magnification factor used in box-counting and is therefore not intrinsic 
to the image. 

In this Letter, we first focus on whether, regardless of these issues, the box-counting curve can be used for authentication. 

\begin{figure}
\begin{center}
\includegraphics[width=0.5\textwidth]{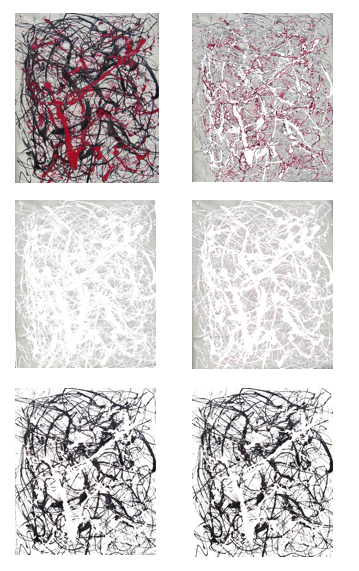}
\end{center}
\caption{{\bf Color Separation} {\em Composition with Red and Black} (top left) and its background layer as obtained with the color separation method of Mureika {\em et al.} 
\cite{mureika} (top right). The background layer for the same painting as obtained by Photoshop (middle left) and RGB Mesh (middle right) methods. The black layer as obtained by Photoshop (bottom left) and RGB Mesh (bottom right) methods. }
\label{fig:colorsep}
\end{figure}

We begin our discussion of image analysis with the problem of color separation. A digital image in RGB mode describes each pixel by three numbers, its R (red), G (green), and B (blue) values, which lie in the range 0 to 255. Naively one might expect individual colors to occupy distinct regions of RGB space,  and to separate a particular color from the rest, one can simply define a median RGB value and collect the pixels that lie within a certain radius. This is precisely the color separation procedure used by Mureika {\em et al.} 
\cite{mureika}.  It is easy to implement but does not work particularly well \cite{colorsepfootnote}, as illustrated in  Fig~\ref{fig:colorsep}. 

As color separation is a delicate task, we developed two independent techniques we call the Photoshop method and the Adaptive RGB Mesh method,  which are both described 
in detail in the accompanying EPAPS Document \cite{epaps}. Results from the two
methods are in excellent agreement \cite{epaps}.

We now embark on a discussion of specific paintings, starting with three famous works  by Pollock. A gallery of all paintings discussed below and their box-counting data
can be found in the accompanying EPAPS document \cite{epaps}.

We chose to analyze {\em Free Form} (1946) because it had previously been included by Taylor {\em et al.} \cite{taylor5} in the list of canonical Pollock paintings from which the fractal authentication criteria were developed \cite{17footnote}.   Surprisingly, we find that {\em Free Form} does not conform perfectly to the fractal authentication criteria; specifically we find that for the composite image $L_T <$1.0 cm \cite{epaps}. 
Although {\em Free Form} does not pass a rigorous application of the fractal authentication criteria, ref \cite{taylor5} describes a more relaxed procedure called force-fitting. Force fitting  imposes the constraint that $L_T >$ 1.0 cm, thereby automatically satisfying authentication criterion (2).  Thus authentication is reduced  to checking that the other criteria are fulfilled. We find that force fitting  {\em Free Form} does not significantly change $D_D$, $D_L$, or $\chi$ and these parameters do remain consistent with fractal authentication criteria. Thus adopting the more relaxed fractal authentication criterion, based solely on fractal analysis one would conclude that {\em Free Form} is indeed authentic. 

\begin{figure}
\scalebox{0.6}{\includegraphics{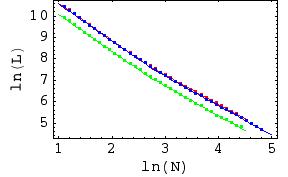}}
\caption{{\bf {\em Untitled} (ca 1950)} (a) Untitled (ca 1950) by Jackson Pollock, Enamel on paper, 28.3 $\times$ cm   150.3 cm, Staatsgalerie Stuttgart \cite{paperbook}. 
The digital image has 1145 $\times$ 5150 pixels. [{\bf Image will be inserted following
publication]}.
(b) Box-counting curves for {\em Untitled} with color separation by the Photoshop method (red) and adaptive RGB mesh (blue) and cropped image (green). Because of the awkward aspect ratio of this painting we were forced to scan it as two sub-images which we then merged. The Ôcropped imageÕ here corresponds to the larger of the two sub-images.}
\label{japanese}
\end{figure}

No such stratagem avails for {\em Untitled} (ca 1950), shown in Fig~\ref{japanese}, which fails the fractal authentication test in a more spectacular way. Force-fitting offers recourse for a painting with $L_T <$ 1.0 cm, and   can be decreased by trimming the range of box sizes; thus anomalous dimensions $D_D > D_L$ is the clearest sign that a painting fails the fractal authenication test.  It is in this respect that Pollock's {\em Untitled} ( ca 1950) fails. {\em The Wooden Horse: Number 10A, 1948} (1948) is another Pollock painting that unambiguously fails to satisfy the fractal authentication test. It has anomalous dimensions in two of six color layers. Thus, based  solely on fractal analysis, one would conclude neither of these paintings (both of which are undisputed Pollock paintings) are authentic Pollocks.  

Interestingly, two of the  25 extant paintings in the cache discovered by Alex Matter resemble {\em Free Form} and {\em Wooden Horse}.  Taylor {\em et al.} analyzed six paintings from the Matter cache and found that none conformed to fractal authentication criteria \cite{abbott}.  We find that the fractal characteristics of the Matter paintings which resemble {\em Free Form} and {\em Wooden Horse} bear a striking similarity to the fractal characteristics of the two analogous Pollock paintings: the painting that resembles {\em Free Form} meets all the fractal authentication criteria provided we force-fit the composite layer, and the painting that resembles {\em Wooden Horse} fails unambiguously by having anomalous dimensions in two out of six layers. 

In \cite{us} we demonstrated that crude drawings such as {\em Untitled 5}  would be mistaken for authentic Pollocks by a computer using fractal analysis for authentication. Although {\em Untitled 5} seriously undermined fractal authentication, it remained to show that actual drip paintings that are unquestionably not by Pollock too can satisfy the fractal authentication criteria; it is this step of the argument we now present. The artists we commissioned, Alexandra Ash and Michael Hallen, studied Pollock's technique and 
rendered nine drip paintings. Of these we have so far analysed two paintings,
chosen for the relative ease with which they could be color separated. As the 
numbers show \cite{epaps}, both paintings satisfy all fractal authentication
criteria comfortably, even without the force-fitting required for 
{\em Free Form}.

We now turn to two problems in fractal mathematics motivated by our analysis of drip paintings but with results of broader significance.  Whether fractal analysis allows authentication of drip paintings has immediate and significant financial implications in the art world, but the mathematical results presented here have more enduring scientific importance. 

 First consider asymptotic scaling behaviour of composites of ideal fractals, e.g., the union of a middle 1/3 Cantor dust (dimension $D_{1/3} = 0.6309\ldots$) and the 3/9 dust (dimension $D_{3/9} = 0.5$)  (see  ref \cite{us} and \cite{epaps}). The 1/3 dust is iterated $2\Lambda_2$  times; the 3/9 dust, $\Lambda_1$  times. The geometry of the union is controlled by $\lambda = \frac{\Lambda_2}{\Lambda_1}$  .These dusts are ideal fractals since the range over which fractal behavior is seen can be made arbitrarily large. In previous work \cite{us} we showed that the union is not fractal because its box counting curve is not a simple power law except in the asymptotic limit of very small box sizes. In that limit, the number of boxes filled with 1/3 dust overwhelms the number filled with 3/9 dust because $D_{1/3} > D_{3/9}$. Thus the union behaves as a fractal with dimension $D_{1/3}$. 

\begin{figure}
\scalebox{0.20}{\includegraphics{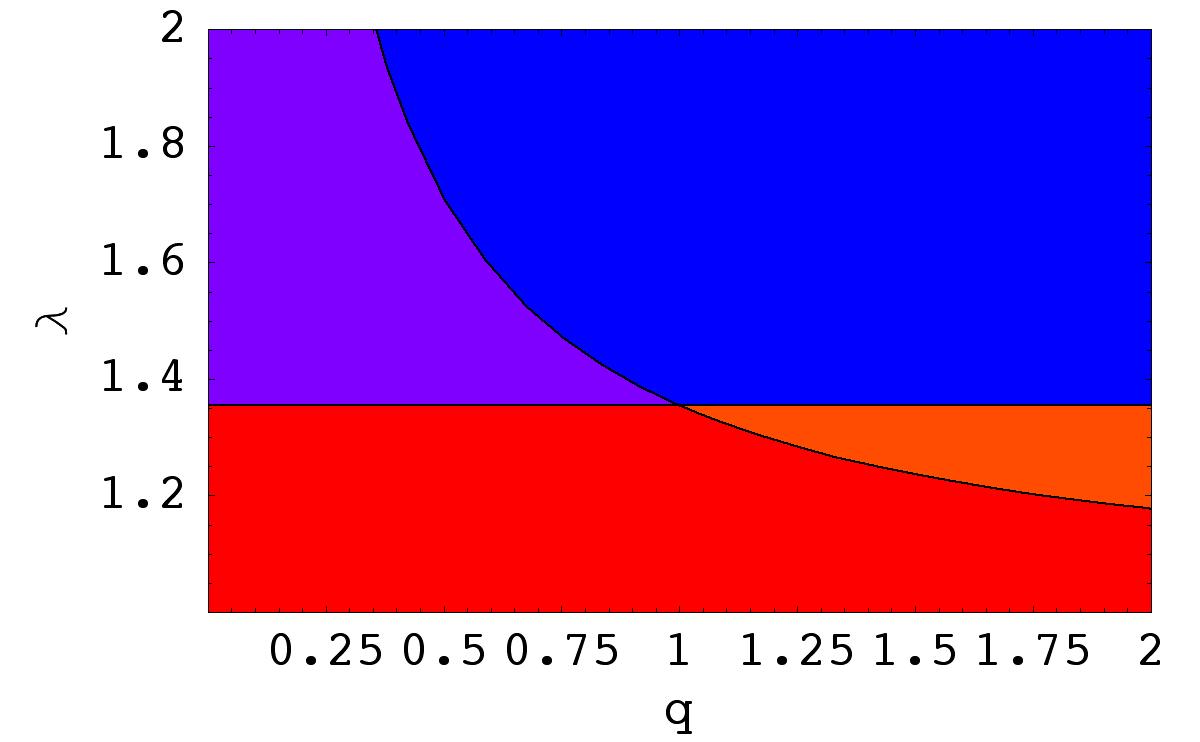}}
\caption{{\bf Phase Diagram for overlapping Cantor Dusts} A 1/3 Cantor dust is iterated  $2\Lambda_2$ times and a 3/9 Cantor dust is iterated $\Lambda_1$  times. The geometry of their union is characterised by a single parameter $\lambda = \frac{\Lambda_2}{\Lambda_1}$. The behaviour of the multifractal spectrum of dimensions $D_q$ of the union is encapsulated by the phase diagram depicted here. In the red phase the 1/3 dust completely dominates and $D_q =D_{1/3} = 0.6309É$ Similarly in the blue phase the 3/9 dust dominates and $D_q  = D_{3/9} = 0.5$. But in the purple region the 3/9 dust dominates on long length scales and the 1/3 on short scales; the effective dimension $D_q$ crosses over from $D_{3/9}$ to $D_{1/3}$. In the orange region the crossover is from 3/9 to 1/3. If we focus on the shortest scales for a fixed $\lambda < \lambda_c = 1.354\ldots$ , $D_q$ jumps from $D_{1/3}$ to $D_{3/9}$ as $q$ is increased. Similarly for fixed $\lambda > \lambda_c$, the jump is from $D_{3/9}$ to $D_{1/3}$. Thus on the shortest scales the union of two fractals is revealed to be a complex multifractal. The vertical line $q = 0$ corresponds to the ordinary fractal dimension. Along this line we find relatively trivial behaviour: the 1/3 dust always dominates and the union behaves asymptotically as a fractal with dimension $D_{1/3}$.}
\label{cantor}
\end{figure}

The full asymptotic complexity of the union is brought out by consideration of the spectrum of multifractal dimensions, $D_q$. On small scales the 1/3 dust determines the fractal dimension, $D_0$, but the 3/9 dust can control $D_q$ for sufficiently large $q$. Fig~\ref{cantor} shows that, except in the trivial instances that one dust overwhelms the other, the union is not scale invariant over the same range as the constituent fractals. Moreover for fixed $\lambda$, varying $q$ can produce a discontinuous jump in the dimension $D_q$. By contrast, for an ideal fractal $D_q$ is independent of  $q$. For generic multifractals $D_q$ varies smoothly with $q$. Discontinuities in $D_q$ are of great interest due to the thermodynamic analogy between multifractal dimensions and phase transitions\cite{ott}. Thus the union is revealed to be a complex multifractal on the shortest length scales; details of the derivation of the 
results quoted here are relegated to \cite{epaps}.

Let us briefly consider the relevance of this analysis to paintings.Taylor {\em et al.} \cite{taylor5}  assert that the composite of two colored layers is a fractal with a dimension that is different from either of its constituent layers. This assertion is mathematically impossible. It is presumably an artefact of fitting power law behaviour to data over a limited range.

The box-counting curve is a Ôstaircase functionÕ, because the number of filled boxes is an integer that increases monotonically as the box size is reduced. Staircase functions abound in mathematics and physics, e.g., $\Pi(x)$, the number of primes less than $x$ \cite{hardy}; $N(E)$, quantum energy levels less than $E$ \cite{mehta}; and $n(s)$, zeta function zeros whose imaginary part lies between zero and $s$ \cite{mehta}. In all these cases there has been a concerted effort to determine a smooth fit through the staircase as well as to characterise the deviations of the staircase from this smooth fit. For example Gauss found a smooth asymptotic fit to $\Pi(x)$. The smooth form of $N(E)$ is basic to statistical mechanics. The deviations from regularity of $N(E)$ show remarkable universal behaviour encapsulated by random matrix theory. The deviation statistics of $n(s)$ are the subject of a conjecture, ancillary to the Riemann hypothesis, due to Dyson and Montgomery. By contrast, hitherto in fractal analysis the focus has been on finding a smooth power law fit to the staircase and extracting the exponent (the fractal dimension); we now study the deviations from regularity. 

For simplicity, instead of the box-counting dimension we analyse a closely related variant \cite{barnsley}, the Ôcovering dimensionÕ. Here the staircase $N(\varepsilon)$  is the number of intervals of size $\varepsilon$  needed to cover the fractal. We first study a unit line segment for which $N(\varepsilon)\approx 1/\varepsilon$, corresponding to a covering dimension of 1. Consider the sequence of interval sizes $\varepsilon_n \equiv \varepsilon_0/C^n $ where $n = 1,\ldots,M$. Here $\varepsilon_0$  is the largest interval size and $C$ is the reduction factor. We define $\chi_n \equiv N(\varepsilon_n) - 1/\varepsilon_n$, the deviation of the exact counting staircase from the smooth fit. It is easy to show that as $M \rightarrow \infty$, $\chi^2$  (the mean value of $\chi_{n}^2$  ) vanishes \cite{epaps}. 
Thus the counting curves of line segments are essentially noiseless. Similar analysis for the middle third Cantor dust shows that the deviates are uniformly distributed over a finite interval and have a non-zero mean square value
\cite{epaps}. Thus the fractal counting curve of the 1/3 dust is intrinsically noisy. Intuitively we can understand that Euclidean objects would be noiseless because the deviations from regular behaviour originate at boundary points. A Euclidean segment has only two boundary points; fractals have an infinity. We conjecture that counting curves are generically noiseless for Euclidean objects whereas they have well defined deviate statistics and mean squared deviation for fractals. Like simple random number sequences \cite{press}, the successive deviates for the 1/3 dust are given by a formula that involves modular arithmetic. The 1/3 dust deviates are highly correlated but it is tantalising to speculate that there might be other fractals whose counting curves could serve as pseudo-random number sources.

Again we briefly consider the implications of these findings for paintings. For fractals  $\chi^2$ is pseudo-random and for generic images it may be $C$ dependent; hence the use of $\chi^2$ as a characteristic of paintings advocated in ref \cite{taylor5} is inappropriate. Taylor {\em et al.} suggest that Pollock's works are high quality fractals because they have small $\chi^2$  values \cite{taylor5}. Our analysis shows that it is
in fact Euclidean objects that have low $\chi^2$. 

In summary, while in \cite{us} we briefly described various shortcomings of fractal analysis as an authentication tool, and identified a number of mathematical inconsistencies lurking in the mere application of fractal analysis to multi-colored drip paintings, we have shown here that (i) amateur artists seeking to emulate Pollock's technique can successfully create paintings which possess the fractal ÒsignatureÓ said to be unique to Pollock; and (ii) even authentic Pollock paintings fail to possess his fractal ÒsignatureÓ. Contrary to Taylor {\em et al.}, we also find at least one Matter painting possesses PollockÕs fractal ÒsignatureÓ.   Finally, our analysis has initiated a study of the statistics of counting staircases, a new topic in fractal mathematics that invites much further exploration. 

Acknowledgements: We thank Alexandra Ash and Michael Hallen for rendering nine drip paintings, and Ellen Landau and David Huse for discussions.

\vspace{3mm}

\begin{center}
{\bf EPAPS AUXILIARY MATERIAL: METHODS}


\vspace{3mm}

{\bf COLOUR SEPARATION}

\end{center}

\vspace{3mm}

\begin{figure}
\begin{center}
\includegraphics[width=0.6\textwidth]{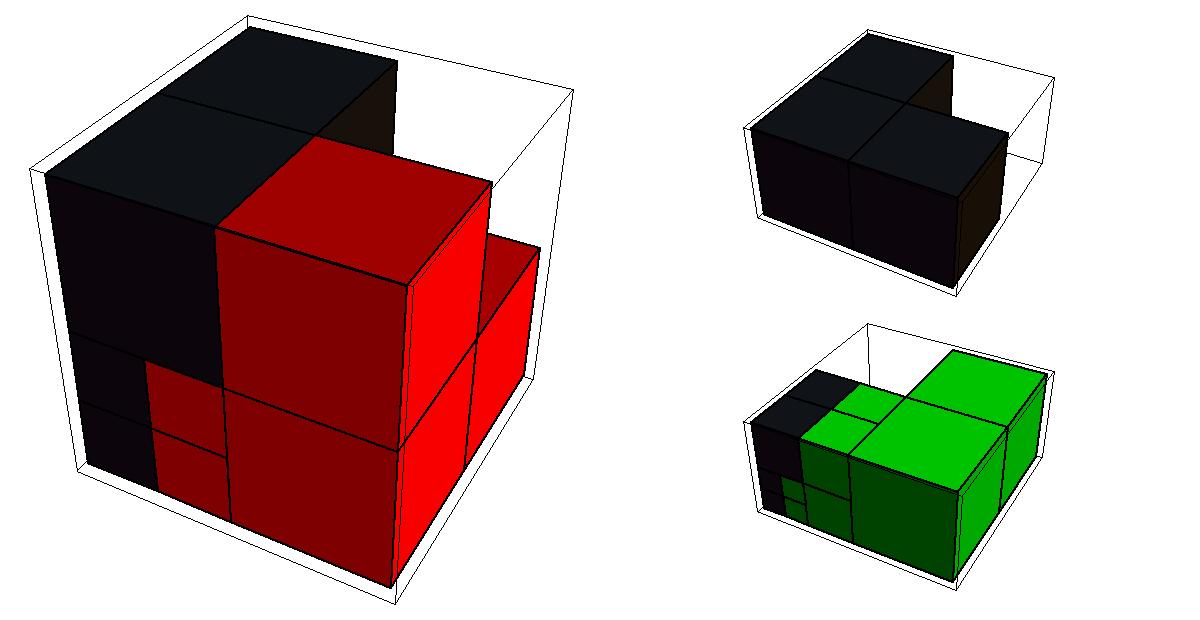}
\end{center}
\caption{The partition of the the RGB cube into red and black regions
for {\em Composition with Red and Black} (left). The blank portion
of the RGB cube is occupied by pixels from the essentially white 
background. The partition of the RGB cube into regions occupied
by black and background pixels for the Matter painting that 
resembles {\em Free Form} is shown to the right. The region of
the RGB cube occupied by background pixels is shown in green; by the 
white layer, left blank. The RGB cube
has been sliced to reveal a cutaway. Note how the different coloured
layers occupy complex interpenetrating regions of RGB colour space.}
\label{fig:interpenetrate}
\end{figure}

In a digital image the colour of each pixel is described by three
numbers each of which lies in the range 0 to 255: its R, G, B values.
Thus colour or RGB space may be viewed as a cube. Each pixel
corresponds to a point inside this cube. Naively one might expect
that pixels of the same colour would clump together in colour space
and pixels of different colours would be well-separated. But in fact
human colour perception is highly non-linear: in 
some regions of colour space a small variation in the RGB values can 
lead to a dramatic change in perceived colour; in other regions, 
large variations in the RGB values correspond to essentially no change
in the perceived colour. Thus different colours form large almost
interpenetrating regions in colour space (see fig \ref{fig:interpenetrate})
and colour separation is therefore a formidable task.

Consider separating a painting such as 
{\em Composition with Red and Black} into its red and black
layers. As discussed above the red pixels and black pixels form 
irregular, interpenetrating swarms in RGB colour-space. The simplest 
approach to colour separation is to place two non-intersecting spheres
inside colour space and to count the pixels inside one as red
and the ones inside the other as black. The location of the
spheres and their size are fitting parameters that can be 
adjusted to ensure that each sphere encloses as many of the
right kind of pixels as possible and as few of the
wrong sort. Typically, though either different colours get mixed,
or parts of a layer get left behind as background residue, 
as illustrated in figure 1 of the paper. 
Thus this method which was used in ref [8], has few virtues besides ease of implementation.
(Parenthetically we note that fig 1 in the paper was generated
by taking the center of the black sphere at (R, G, B) $= (41, 40, 41)$
and the red sphere at (161, 15, 34). The radius of both spheres was
taken to be 40. The centers were chosen by using the median (R,G,B)
values, as computed using Photoshop, and the radius was optimised
by trial and error). 

In the Photoshop method the user selects a pixel using a tool
called magic wand. All pixels within a narrow user-controlled
RGB range (the ``tolerance'') of the selected pixel are then highlighted 
by magic wand. If they are all from the desired layer they are  
transferred to a blank background. The user continues to select pixels
in this way 
until the entire layer has been transferred. By judicious selection of
pixels and use of a narrow tolerance accidental selection of pixels from
a different layer can be minimised. Schematically the contrast between
the simple method of the previous paragraph and the Photoshop method
is that the former attempts to enclose the swarm of red pixels within
a single well-chosen sphere; the Photoshop method covers it with a 
series of small spheres and can therefore accomodate the large extent
and the irregular shape.

In the adaptive RGB mesh approach the colour cube is divided up
into eight smaller cubes which are designated with a binary address.
011 for example refers to the sub-cube where 
$0 < {\rm R} < 127, 128 < {\rm G} < 255$ and $128 < {\rm B} < 255$; 101, 
where $128 < {\rm R} < 255, 0 < {\rm G} < 127$ and $128 < {\rm B} < 255$; 
and so forth.
The user then examines an image in which only pixels from a
single subcube are displayed with their true colours; the other 
pixels are displayed as white or another user selected uniform
background colour. If this image is seen to contain pixels
from a single layer the user records that the entire subcube
contains only pixels of one colour. For example, for {\em Composition
with Red and Black} it was found that 001, 010 and 011 contained only
black pixels; 100, 101 and 110 contained only red; 111 contained only
background. But if a subcube contains more than one colour, as does
000, it is further subdivided into eight cubes, and the process is
reiterated as many times as needed. {\em Composition in Red and Black}
for example required three levels of subdivision. The binary address
notation for higher level subdivisions works in the same way.
For example the level two sub-cube with binary address 01-10-11 refers
to the region where $64 < {\rm R} < 127, 128 < {\rm G} < 191$ and 
$192 < {\rm B} < 255$. 

The two methods have complementary strengths. The Photoshop method,
if skillfully deployed, is faster. Also it allows the user to make
better use of spatial information when deciding which pixels to 
include in a given layer. The adaptive RGB mesh on the other
allows for systematic refinement and provides a detailed 
characterisation of the way in which colour space is partitioned 
into the layers of the drip painting. The code we developed to
implement RGB mesh included a number of features including the
ability to count the pixels in each sub-cube to determine their
relative importance and the ability to regroup and consolidate
pixels spatially for easier viewing. In principle the program
could be refined to make fuller use of spatial information
but in practice we found that results adequate for our needs
could be obtained without this feature.

\vspace{3mm}

\begin{center}

{\bf CONSTRUCTION OF CANTOR DUSTS}

\end{center}

\vspace{3mm}

The 1/3 dust [11, 14] is constructed by starting with a unit line segment and 
removing the middle third. At the second iteration the middle third of the
two surviving segments is removed; at the third iteration, the middle third
of the four surviving segments. After $\Lambda$ iterations the set
consists of $2^{\Lambda}$ segments each of length $1/3^{\Lambda}$.
This set behaves as a fractal on length scales between $1$ and $1/3^{\Lambda}$.
By increasing $\Lambda$ we can make this range arbitrarily large, thus approaching
a mathematical fractal. 

There is a useful way to label segments using base
3 numbers. At the first iteration we label the left, middle 
and right thirds of the initial unit segment as 0, 1 and 2 
respectively. We then label the left, middle and right thirds of
the left third 00, 01, and 02. Proceeding in this way we see that
after the second iteration the four segments that survive have the labels
00, 02, 20 and 22. After $\Lambda$ iterations there are $3^{\Lambda}$
segments. The address of each segment is a base 3 number of $\Lambda$ digits.
The segments that survive are the ones that have only 0s and 2s in their base three
addresses.

The 3/9 dust is constructed by starting with a unit line segment, dividing it
into nine equal segments, and retaining only the first, fifth and ninth segments [7]. 
At the second iteration the surviving segments are divided up into ninths and
in each case only the first, fifth and ninth subsegments are retained. After 
$\Lambda$ iterations the set consists of $3^{\Lambda}$ segments each of
length $1/9^{\Lambda}$. Segments in the 3/9 dust can be labelled by using
base 9 numbers. For example, after the first iteration the segments that are retained are 
labelled 0, 4 and 8. After $\Lambda$ iterations the segments will have labels that
consist of base 9 numbers of $\Lambda$ digits. The segments that belong to the 3/9 dust
will have only 0s, 4s and 8s  in their address.

Another point of view on the 1/3 dust  is that it is obtained by taking the
unit segment, dividing it up into ninths, and retaining the first, third, seventh
and ninth segments. From this point of view, after $2 \Lambda$ iterations
the dust consists of $4^{\Lambda}$ segments each of length $1/9^{\Lambda}$.
From this perspective segments can be labelled by base 9 numbers. For example,
after the first iteration the four surviving segments are 0, 2, 6 and 8. After 
$2 \Lambda$ iterations the segments will have labels that consist of base nine numbers of
$\Lambda$ digits. The ones that belong to the 1/3 dust will have only 0s, 2s, 6s and
8s in their address. This viewpoint on the 1/3 dust is useful when studying its 
union and intersection with the 3/9 dust.

\vspace{3mm}

\begin{center}

{\bf ASYMPTOTIC ANALYSIS OF OVERLAPPING FRACTALS}

\end{center}

\vspace{3mm}


\noindent
{\bf (a) Two Cantor Dusts:} 
First we summarize our results, then we outline
their derivation.
Consider the union of a 1/3 Cantor dust iterated $2 \Lambda_2$ times 
and a 3/9 dust iterated $\Lambda_1$ times with $\lambda \equiv \Lambda_2 / \Lambda_1
\geq 1$.
This construction results in $4^{\Lambda_2}$ segments each of length $1/9^{\Lambda_2}$
and $ 3^{\Lambda_1} - 2^{\Lambda_1}$ segments each of length $1/9^{\Lambda_1}$.
The total length of the resulting object is therefore
\begin{equation}
l_{{\rm tot}} = 4^{\Lambda_2} \frac{1}{9^{\Lambda_2}} + 
(3^{\Lambda_1} - 2^{\Lambda_1}) \frac{1}{9^{\Lambda_1}}.
\label{eq:ltot}
\end{equation}
We suppose that the resulting object has unit mass and that the density
(mass per unit length) is uniform and hence equal to $1/l_{{\rm tot}}$.

We now cover this object with boxes of size $\epsilon = 1/9^{\mu}$ where
$ 1 \leq \mu \leq \Lambda_1$. It turns out that there are three classes of filled
boxes. The number  of boxes of each class and the mass contained within each
box are given by
\begin{eqnarray}
n_A & = & 2^{\mu}, 
\nonumber \\
m_A & = & \left[
4^{\Lambda_2 - \mu}/9^{\Lambda_2} + 
(3^{\Lambda_1 - \mu} - 2^{\Lambda_1 - \mu} )/9^{\Lambda_1} \right]/l_{{\rm tot}}; 
\nonumber \\
n_B & = & 4^{\mu} - 2^{\mu}, 
\nonumber \\
m_B & = & 
\left[4^{\Lambda_2 - \mu} /9^{\Lambda_2} \right]/l_{{\rm tot}};  
\nonumber \\
n_C & = & 3^{\mu} - 2^{\mu},
\nonumber \\
m_C & = & 
\left[3^{\Lambda_1 - \mu} /9^{\Lambda_1} \right]/l_{{\rm tot}}. 
\label{tab:mass}
\end{eqnarray}



Instead of the number of filled boxes of size $\epsilon$, in multifractal analysis$^{13}$
we consider the generalised box count $N(q, \epsilon)$ defined as
\begin{equation}
\ln N(q, \epsilon) = \frac{1}{1-q} \ln \sum_{i} m_i^q.
\label{eq:generalboxcount}
\end{equation}
Here $q$ is a continuous real parameter, $m_i$ is the mass in the $i^{{\rm th}}$ 
occupied box, and the sum is over all filled boxes. For multifractals it is expected that
$N(q, \epsilon)$ will vary as a power of box size $\epsilon$. Thus we may extract
the multifractal spectrum of dimensions by computing
\begin{equation}
D_q =  \lim_{\epsilon \rightarrow 0} \frac{\ln N(q, \epsilon)}{\ln (1/\epsilon)}.
\label{eq:multidimension}
\end{equation}
Evidently $N(q,\epsilon)$ is 
equal to the number of filled boxes for $q = 0$ and $D_0$ is just the ordinary
fractal dimension.
Making use of eq (\ref{tab:mass}), it follows that for the overlapping Cantor dusts
\begin{equation}
N(q, \epsilon) = \frac{1}{1-q} \ln ( n_A m_A^q + n_B m_B^q + n_C m_C^q ).
\label{eq:twodust}
\end{equation}
Eq (\ref{eq:twodust}) is the central result of this section. It allows us to
plot the generalised box counting curve of the union of the Cantor dusts as
well as to determine the asymptotic behaviour.

The phase diagram for the overlapping Cantor dusts can be constructed by
straightforward asymptotic analysis of eq (\ref{eq:twodust}). It is easy to argue
that class B and class C boxes generally dominate the contribution of class A
to the generalised box count. When class B dominates class C the generalised
box count approaches that of a pure 1/3 dust; when class C dominates class B,
that of a pure 3/9 dust. First consider the case $0 < q < 1$. In this case
for $1 < \lambda < \lambda_c $ the 1/3 dust dominates; for $ \lambda > \lambda^+$
the 3/9 dust dominates. For $ \lambda_c < \lambda < \lambda^+$ there is a crossover:
the 3/9 dust dominates for $\mu < \mu_c$ and the 1/3 dust dominates for $ \mu > \mu_c$.
Here $\lambda_1 = \ln 3/\ln(9/4) = 1.354\ldots$, $ \lambda^+$ is the curve defined by
$ \lambda^+ = 1 + (\lambda_c - 1)/q$, and $\mu_c$ is given by
\begin{equation}
\mu_c = \Lambda_1 \frac{q}{1-q} \frac{\left[ \lambda \ln (9/4) - \ln 3 \right]}{\ln(4/3)}.
\label{eq:muc}
\end{equation}
Note that $\mu_c/\Lambda_1$ varies from zero to one as $\lambda$ varies from $\lambda_c$ to
$\lambda^+$.

Similarly for $ q > 1$ we find that for $ \lambda > \lambda_c$ the 3/9 dust dominates.
For $ \lambda^+ > \lambda > 1$ the 1/3 dust dominates. For $ \lambda^+ > \lambda >
\lambda_c$ there is a crossover: the 1/3 dust dominates for $\mu < \mu_c$ and the
3/9 for $\mu > \mu_c$ where $\mu_c$ is given by eq (\ref{eq:muc}). 

Now we outline the derivation of these results. For simplicity we only consider the union of
two Cantor dusts with $\Lambda_2 = \Lambda_1 = \Lambda$; the generalisation is straightforward. 
We adopt the notation that 
$n = 0$ or $8$, $m = 2$ or $6$,  $\nu = 0, 4,$ or $8$ and $N=0,2,6$ or 8. 
After the first iteration of the 3/9 dust and the second iteration of the 1/3 dust 
the base nine addresses of the surviving 
segments are 0, 2, 4, 6 and 8. In terms of the notation above the
addresses are $n_1$, $m_1$ or 4. At the next iteration the possible addresses 
are: $n_1 n_2$; $n_1 m_2$, $m_1 N_2$; $n_1$ 4 or 4 $\nu_2$. By working out a few
iterations with this notation it is easy to deduce that once an $m$ appears in an
address the subsequent integers in the address must be $N$'s. Once a 4 appears,
the subsequent integers are $\nu$'s. On the other hand, an $n$ can be followed by $n$, $m$
or 4. Thus after $\Lambda$ iterations the segments fall into three classes
with binary addresses that (A) are composed entirely of $n$'s, (B) include 
at least one $m$ and (C) include at least one $4$. By induction we can show
that $n_A = 2^{\Lambda}$, $n_B = 4^{\Lambda} - 2^{\Lambda}$
and $n_C = 3^{\Lambda} - 2^{\Lambda}$. Since all segments are of length
$1/9^{\Lambda}$ we can easily write down an expression for the total length
of the object $l_{{\rm tot}}$ which is in agreement with eq (\ref{eq:ltot})

Now suppose we cover the object with boxes of size $1/9^{\mu}$ where $\mu < \Lambda$.
The boxes themselves fall into the three classes A, B, and C depending on their base nine
addresses. A class A box has only $n$'s in its base nine address. A segment that lies in a
class A box has the same sequence of $n$'s for the first $\mu$ digits of its base nine
address. The remaining $\Lambda - \mu$ digits may be $n$'s, $m$'s or 4s. Thus the
segments contained in a class A box effectively constitute the union of a 1/3 and 3/9 dust
that is iterated $\Lambda - \mu$ times. Therefore it contains 
$4^{\Lambda - \mu} + 3^{\Lambda - \mu} - 2^{\Lambda - \mu}$ segments.
Moreover, since the address of a class $A$ box is a string of $n$'s of length $\mu$, evidently 
there are $2^{\mu}$ class A boxes. In the same way we can show that there are 
$4^{\mu} - 2^{\mu}$ boxes of class B and $3^{\mu} - 2^{\mu}$ boxes of class C. The segments
in a class B box have a base nine address that coincides with that of the box for the first
$\mu$ places. According to the rules above this is followed by a string of $N$'s
of length $\Lambda - \mu$. Thus a class B box contains $4^{\Lambda - \mu}$ segments.
Similarly a segment in a class C box has a base nine binary address that coincides with
that of the box for the first $\mu$ places. Thereafter it is a string of $\nu$'s of length
$\Lambda - \mu$. Thus a class C box contains $3^{\Lambda - \mu}$ segments.
With this we have derived all the information in eq (\ref{tab:mass}) for the special
case $\Lambda_1 = \Lambda_2 = \Lambda$. The derivation is readily generalised 
to the case $\Lambda_1 \neq \Lambda_2$. 

Note also that for this special case 
we have also
shown that the number of filled boxes of size $1/9^{\mu}$ is $4^{\mu} + 3^{\mu} - 2^{\mu}$.
This is the result plotted in fig 1 of ref [7] to illustrate that the box counting curve for the
union of two fractals is not a simple power law.

\vspace{2mm}

\noindent
{\bf (b) Euclidean Island in sea of Cantor Dust:}
We now briefly outline a second soluble example of the union of two objects with
distinct fractal dimensions. The set we consider consists of a 1/3 dust that is initially iterated
$\Lambda_1$ times. The segment to the extreme right (the Euclidean island) is now 
left fixed. The other surviving segments are further subdivided a further $\Lambda_2$ times
to form a fine 1/3 dust.
The Euclidean island has a length $1/3^{\Lambda_1}$.
The dust consists of $2^{\Lambda_1 + \Lambda_2} - 2^{\Lambda_2}$ segments each of 
length $1/(3^{\Lambda_1 + \Lambda_2})$. The total length of the object is therefore
\begin{equation}
l_{{\rm tot}} = \frac{1}{3^{\Lambda_1}} + ( 2^{\Lambda_1 + \Lambda_2} - 2^{\Lambda_2} )
\frac{1}{3^{\Lambda_1 + \Lambda_2}}.
\label{eq:ltotdustisland}
\end{equation}

Let us now cover the object with boxes of size $\epsilon = 1/(3^{\Lambda_1 + \mu})$
where $ 0 \leq \mu \leq \Lambda_2$. The occupied boxes fall into class (A) that cover
dust and class (B) that cover the Euclidean island. 
It is easy to see that
\begin{eqnarray} 
n_A = 2^{\Lambda_1 + \mu} - 2^{\mu}, & &
m_A = \frac{1}{l_{{\rm tot}}} 2^{\Lambda_2 - \mu} \frac{1}{3^{\Lambda_1 + \Lambda_2}};
\nonumber \\
n_B = 3^{\mu}, & & m_B = \frac{1}{l_{{\rm tot}}} \frac{1}{3^{\Lambda_1 + \mu}}.
\label{eq:table}
\end{eqnarray}
Thus the generalised box count defined in eq (\ref{eq:generalboxcount}) is given
by
\begin{equation}
\ln N(q, \epsilon) = \frac{1}{1-q} \ln( n_A m_A^q + n_B m_B^q ).
\label{eq:multiisland}
\end{equation}
Eq (\ref{eq:multiisland}) is the main result of this subsection. 
It allows us to plot the generalised box counting curve for this composite
object as well as to analyse its asymptotic behaviour. 

\begin{figure}
\begin{center}
\includegraphics[width=0.45\textwidth]{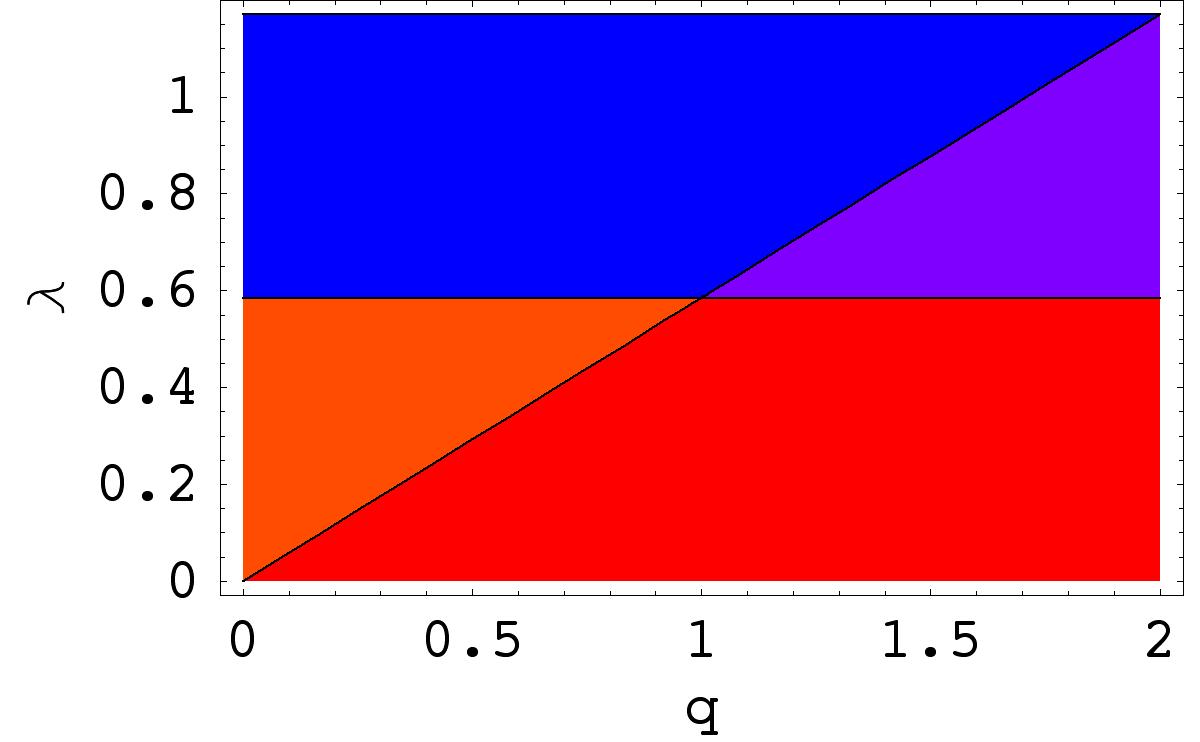}
\end{center}
\caption{{\bf Phase Diagram for Euclidean segment and Cantor Dust: } 
The geometry of the union of the two sets is controlled by a single parameter
$\lambda = \Lambda_1/\Lambda_2$ where $\Lambda_1$ is the scale of the
island and $\Lambda_1 + \Lambda_2$ of the dust as explained in the text.
The diagram shows the behaviour of the multifractal spectrum of dimensions
$D_q$ as a function of $\lambda$ and $q$. In the red phase the island dominates
completely and $D_q = 1$. Similarly in the blue phase the 1/3 dust dominates and
$D_q = D_{1/3}$. But in the purple region the island dominates on long scales while
the dust dominates on short scales; the effective dimension $D_q$ crosses over from
1 to $D_{1/3}$. In the orange region the dust dominates on long length scales and the
island on short scales; the effective dimension $D_q$ crosses over from 
$D_{1/3}$ to 1. If we fix $\lambda$, $D_q$ jumps discontinuously as a function
of $q$ over an appropriate range of length scales. Thus the union of the two sets
is a complex multifractal on these scales.}
\label{fig:euclphase}
\end{figure}

The phase diagram implied by eqs (\ref{eq:table}) and (\ref{eq:multiisland})
can be obtained by straightforward asymptotic analysis. The geometry of the 
union of the island and dust is controlled by a single parameter $\lambda \equiv
\Lambda_1/\Lambda_2$.  First consider 
$ 0 \leq q < 1$.  It is convenient to define $\lambda_0 = \ln(3/2)/\ln2 = 0.58496\ldots$.
Now for $ \lambda < q \lambda_0$ the island dominates and the dimensions
$D_q = 1$. For  $\lambda > \lambda_0$ the dust dominates and the dimensions
$D_q = D_{1/3}$, the fractal dimension of the 1/3 dust. 
For $ q \lambda_0 < \lambda < \lambda_0$ the dust dominates for $\mu < \mu_0$,
the island for $\mu > \mu_0$; the slope of the generalised boxcounting curve 
crosses over from $D_{1/3}$ to 1. Here the crossover scale $\mu_0$ is given by
\begin{equation}
\mu_0 = \Lambda_2 \frac{1}{1-q} \frac{ \lambda - q \lambda_0 }{ \lambda_0 }
\label{eq:euclislandcross}
\end{equation}
Now consider $ q > 1$. In this case for $\lambda < \lambda_0$ the island 
dominates and the dimensions $D_q = 1$. For $\lambda > q \lambda_0$ 
the dust dominates and $D_q = D_{1/3}$. For $ \lambda_0 < \lambda < q \lambda_0$
the island dominates for $\mu < \mu_0$ and the dust for $\mu > \mu_0$; the slope of
the generalised box-counting curve crosses over from 1 to $D_{1/3}$. This information
is encapsulated in the phase diagram shown in fig \ref{fig:euclphase}.

\vspace{2mm}

\begin{center}

{\bf COUNTING STATISTICS}

\end{center}

\vspace{2mm}

\noindent
{\bf (a) Line segment:} $N(\epsilon)$ is the smallest number of segments required
to cover a unit line segment [14]. Evidently
\begin{equation}
N(\epsilon) = {\rm Ce} \left( \frac{1}{\epsilon} \right)
\label{eq:segstair}
\end{equation}
where ${\rm Ce}(x)$ is the smallest integer that is greater than or equal to $x$. 
In other words Ce rounds up its argument. $N_{{\rm smooth}} (\epsilon) = 1/\epsilon$
is a smooth fit through this counting staircase. From the definition
\begin{equation}
\chi(\epsilon) = \ln N(\epsilon) - \ln N_{{\rm smooth}}(\epsilon) 
\label{eq:chie}
\end{equation}
and eq (\ref{eq:segstair}) it is easy to see that for small $\epsilon$ 
\begin{equation}
\chi(\epsilon) = \epsilon \Delta(\epsilon).
\label{eq:segchi}
\end{equation}
Here $\Delta(\epsilon) = {\rm Ce}(1/\epsilon) - 1/\epsilon$ lies between zero and
one since it represents the amount of round up. 

We now sample the counting curve at a sequence of interval sizes $\epsilon_n = 
\epsilon_0/C^n$ where $C > 1$ is a reduction factor, $\epsilon_0$ is the largest 
interval size considered and $n = 0, 1, 2, \ldots, M-1$. The quantities of interest 
are $\chi_n = \chi(\epsilon_n)$, the deviations of the counting curve from the
smooth fit at the sampled interval sizes. As usual the mean square deviation is
given by
\begin{equation}
\chi^2 = \frac{1}{M} \sum_{n=0}^{M-1} \chi_n^2.
\label{eq:chisquare}
\end{equation}
It is easy to derive the bound
\begin{equation}
\chi^2 \leq \frac{\epsilon_0^2}{M} \frac{1}{1 - 1/C^2}
\label{eq:segchisquare}
\end{equation}
revealing that $\chi^2 \rightarrow 0$ as $M \rightarrow \infty$. Thus Euclidean staircases
are essentially noiseless. 

\vspace{2mm}

\noindent
{\bf (b) Cantor dust:}
The analysis of the middle third Cantor dust proceeds exactly in parallel to the 
Euclidean line segment. The exact staircase function is given by
\begin{eqnarray}
N(\epsilon) & = & 2^{n(\epsilon)} 
\nonumber \\
{\rm where} \hspace{2mm} n(\epsilon) & = & {\rm Ce} \left[ 
\frac{\ln(1/\epsilon)}{\ln 3} \right].
\label{eq:cantorstair}
\end{eqnarray}
$N_{{\rm smooth}} = 1/\epsilon^{D_{1/3}}$ is a smooth fit through this staircase
where $D_{1/3} = \ln2/\ln3$ is the fractal dimension of the middle third dust.
From the definition of $\chi(\epsilon)$, eq (\ref{eq:chie}), and the exact staircase,
eq (\ref{eq:cantorstair}), it follows that
\begin{equation}
\chi(\epsilon) = \ln 2 \left(
{\rm Ce} \left[ \frac{ \ln(1/\epsilon)}{\ln 3} \right] - 
\frac{ \ln(1/\epsilon)}{\ln 3} \right).
\label{eq:cantorchi}
\end{equation}
By sampling the counting curve in the usual way we obtain the
sequence of deviates
\begin{equation}
\chi_n = \ln 2 \left(
{\rm Ce} 
\left[ \frac{ n \ln C }{ \ln 3 } - \frac{ \ln \epsilon_0 }{ \ln 3 } \right] - 
\left[ \frac{ n \ln C }{ \ln 3 } - \frac{ \ln \epsilon_0 }{ \ln 3 } \right]  \right).
\label{eq:cantorchin}
\end{equation}
This leads us to contemplate expressions of the form
\begin{equation}
\xi_n = {\rm Ce} ( a n + b ) - (a n + b).
\label{eq:xin}
\end{equation}
Let us make the rational choice $a = p/q$ where, without further loss of generality,
we may take $p$ and $q$ to be coprime and $p < q$. From eq (\ref{eq:xin}) it is 
easy to see that $\xi_n$ are a uniformly spaced sequence of $q$ points on the unit 
interval. Thus the $\chi_n$'s are seen to be uniformly distributed and highly correlated.

With suitable rescaling eq (\ref{eq:xin}) may be written in the form
$I_{n+1} = (\alpha n + \gamma) \hspace{1mm} ({\rm mod} \hspace{2mm} m)$ 
where $I_n = q \xi_n$ and 
$\alpha$, $\gamma$  and $m$ are suitable constants. This form is very
close to a common random number generator $I_{n+1} = (\alpha I_n + \gamma) 
\hspace{1mm} ({\rm mod} \hspace{2mm} m)$ [15]. 
An interesting question raised by this similarity, 
one that we leave open, is whether there is a deterministic fractal whose counting staircase 
generates an acceptable pseudo-random sequence. 

\vspace{3mm}

\begin{center}

{\bf EPAPS AUXILIARY MATERIAL: DATA}

\vspace{3mm}

{\bf ANALYSIS OF DRIP PAINTINGS}

\end{center}

\vspace{2mm}

\noindent
{\bf Notes:}  Results obtained using the Photoshop colour separation method are
marked PS; using adaptive RGB mesh, marked RGB. Known Pollock works are
identified by name and their serial number in {\em Jackson Pollock: A Catalogue
Raisonn\'{e} of Paintings, Drawings and Other Works} by F.V. O'Connor and E.V. Thaw
(Yale University Press, New Haven, 1978), hereafter abbreviated JPCR. 
For some paintings 
the digital image was cropped slightly before analysis. In these cases we 
give the dimensions of the cropped image used in each analysis. For layers
that were force fit the results are marked ff; see below for a precise description
of force-fitting. The quantity that is directly inferred
from the data is $\ln L_T$ but we present $L_T$ values since it is more easy to
visualise. However it should be kept in mind that by exponentiating the measured
quantity we are amplifying the errors. Thus our two methods agree much better
on the location of the box-counting transition than this tabulation of $L_T$ values 
may suggest. For each painting the results are briefly summarised at the end.

\vspace{2mm}

\noindent
{\bf Force-fitting:} 
Ref [7] describes a more relaxed procedure called force-fitting that can be applied
to paintings that fail to pass the strict authentication test. Force-fitting imposes the 
constraint that $L_T > 1.0$ cm, thereby automatically satisfying authentication criterion (2). 
Thus authentication is reduced to checking that the other criteria are fulfilled. 

\vspace{2mm}

\noindent
{\bf 1. Untitled 1:} (Poured media on board). Matter painting that resembles Pollock's
{\em Free Form}.

\vspace{2mm}

\noindent
Physical Dimensions: 35.6 $\times$ 44.5 cm

\noindent
Digital Image: 2020 $\times$ 2520 (PS black) 1994 $\times$ 2487 (PS white)
1986 $\times$ 2502 (RGB)

\noindent
Range of box sizes:
0.2 cm to 3.1 cm (PS) 0.2 cm to 3.1 cm (RGB)

\vspace{2mm}

\noindent
{\em Black Layer:}

\noindent
Area of layer: 43.0 \% (PS) 38.46 \% (RGB) 

\noindent
Noise: 0.023 (PS) 0.017 (RGB) 0.020 (RGB ff)

\noindent
$L_{T}$: 1.0 cm (PS) 0.8 cm (RGB) 1.0 cm (RGB ff)

\noindent
Lower dimension: 1.41 (PS) 1.71 (RGB) 1.73 (RGB ff)

\noindent
Upper dimension: 1.98 (PS) 1.98 (RGB) 2.00 (RGB ff)

\vspace{2mm}

\noindent
{\em White Layer:} 

\noindent
Area of Layer: 13.9 \% (PS) 10.86 \% (RGB) 

\noindent
Noise: 0.024 (PS) 0.023 (RGB) 0.025 (PS ff) 0.024 (RGB ff)

\noindent
$L_{T}$: 0.85 cm (PS) 0.95 cm (RGB) 1.0 cm (PS ff) 1.0 cm (RGB ff)

\noindent
Lower dimension: 1.47 (PS) 1.35 (RGB) 1.49 (PS ff) 1.36 (RGB ff)

\noindent
Upper dimension: 1.97 (PS) 1.93 (RGB) 2.00 (PS ff) 1.95 (RGB ff)

\vspace{2mm}

\noindent
{\em Composite Layer:}

\noindent
Noise: 0.018 (PS) 0.017 (RGB) 0.021 (PS ff) 0.021 (RGB ff)

\noindent
$L_T$: 0.58 cm (PS) 0.59 cm (RGB) 1.0 cm (PS ff) 1.0 cm (RGB ff)

\noindent
Lower Dimension: 1.86 (PS) 1.82 (RGB) 1.90 (PS ff) 1.86 (RGB ff)
 
\noindent
Upper Dimension: 2.02 (PS) 2.02 (RGB) 2.07 (PS ff) 2.04 (RGB ff)

\vspace{2mm}

\noindent
{\em RGB Colour partition:}

\noindent
White = 111, 010

\noindent
Black = 001, 011, 101, 
\newline
00-00-01, 00-01-00, 00-01-01,
\newline
000-000-000, 000-000-001, 000-001-000, 000-001-001

\noindent
Brown = 100, 110, 
\newline
01-00-00, 01-00-01, 01-01-00, 01-01-01,
\newline
001-000-000, 001-000-001, 001-001-000, 001-001-001

\vspace{2mm}

\noindent
{\bf Summary:} Satisfies fractal authentication criteria if the more lax force-fitting procedure
is allowed.

\vspace{3mm}

\noindent
{\bf 2. Untitled 14:} (Poured media on board). Matter painting that resembles 
Pollock's {\em Wooden Horse}.

\vspace{2mm}
\noindent
Physical Dimensions: 31.7 cm $\times$ 46.4 cm. 

\noindent
Digital Image: 1764 $\times$ 2568 pixels

\noindent
Range of box sizes: 0.2 cm to 3.1 cm (PS)

\vspace{2mm}

\noindent
{\em Red Layer:}

\noindent
Area of Layer: 10.0 \% (PS)

\noindent
Noise: 0.014 (PS)

\noindent
$L_T:$ 1.4 cm (PS)

\noindent
Lower Dimension: 1.36 (PS)

\noindent
Upper Dimension: 1.69 (PS)

\vspace{2mm}

\noindent
{\em Orange Layer:}

\noindent
Area of Layer: 1.7 \% (PS)

\noindent
Noise: 0.024 (PS)

\noindent
$L_T:$ 0.5 cm (PS)

\noindent
Lower Dimension: 1.28 (PS)

\noindent
Upper Dimension: 0.91 (PS)

\vspace{2mm}

\noindent
{\em Yellow Layer:}

\noindent
Area of Layer: 5.1 \% (PS)

\noindent
Noise: 0.018 (PS)

\noindent
$L_T:$ 1.2 cm (PS)

\noindent
Lower Dimension: 1.08 (PS)

\noindent
Upper Dimension: 1.69 (PS)

\vspace{2mm}

\noindent
{\em Blue Layer:}

\noindent
Area of Layer:  4.2 \% (PS)

\noindent
Noise: 0.023 (PS)

\noindent
$L_T:$ 1.0 cm (PS)

\noindent
Lower Dimension: 1.05 (PS)

\noindent
Upper Dimension: 1.67 (PS)

\vspace{2mm}

\noindent
{\em Black Layer:}

\noindent
Area of Layer: 11.6 \% (PS)

\noindent
Noise: 0.014 (PS)

\noindent
$L_T:$ 1.2 cm (PS)

\noindent
Lower Dimension: 1.36 (PS)

\noindent
Upper Dimension: 1.71 (PS)

\vspace{2mm}

\noindent
{\em White Layer:}

\noindent
Area of Layer: 11.0 \% (PS)

\noindent
Noise: 0.010 (PS)

\noindent
$L_T:$ 2.0 cm (PS)

\noindent
Lower Dimension: 2.02 (PS)

\noindent
Upper Dimension: 1.81 (PS)

\vspace{2mm}

\noindent
{\bf Summary:} Does not satisfy fractal authentication criteria.
Exhibits anomalous dimensions in orange and white layers. 

\vspace{3mm}

\noindent
{\bf 3. Free-Form} Pollock (1946) [JPCR 165] (Oil on canvas).

\vspace{2mm}

\noindent
Physical Dimensions: 48.9 cm $\times$ 35.5 cm

\noindent
Digital Image: 4550 $\times$ 3253 (PS Black), 4576 $\times$ 3288 (PS 
White); 2277 $\times$ 1629 pixels (RGB)

\noindent
Range of box sizes: 0.1 cm to 4.5 cm (PS) 0.2 cm to 4.2 cm (RGB)

\vspace{2mm}

\noindent
{\em NB:} the image was coarsened before RGB analysis to speed up 
computations. In principle this should have no effect on the fractal analysis
since the smallest box sizes exceed the coarsening scale. This expectation
is borne out by the concordance between the PS and RGB results.  

\vspace{2mm}

\noindent
{\em Black Layer:}

\noindent
Area of Layer: 56.0 \% (PS) 54.39 \% (RGB)

\noindent
Noise: 0.015 (PS) 0.013 (RGB)

\noindent
$L_T$: 1.0 cm (PS) 1.0 cm (RGB)

\noindent
Lower dimension: 1.79 (PS) 1.85 (RGB)

\noindent
Upper dimension: 2.05 (PS) 2.02 (RGB)

\vspace{2mm}

\noindent
{\em White Layer:}

\noindent
Area of Layer: 20.7 \% (PS) 23.48 \% (RGB)

\noindent
Noise: 0.019 (PS) 0.019 (RGB) 0.021 (PS ff) 

\noindent
$L_T$: 0.9 cm (PS) 1.0 cm (RGB) 1.0 cm (PS ff)

\noindent
Lower dimension: 1.57 (PS) 1.60 (RGB) 1.58 (PS ff)

\noindent
Upper dimension: 2.04 (PS) 2.00 (RGB) 2.06 (PS ff)

\vspace{2mm}

\noindent
{\em Composite Layer:}

\noindent
Noise: 0.013 (PS) 0.012 (RGB) 0.012 (RGB ff) 0.015 (PS ff)

\noindent
$L_T$: 0.8 cm (PS) 0.65 cm (RGB) 1.2 cm (RGB ff) 1.1 cm (PS ff)

\noindent
Lower dimension: 1.93 (PS) 1.94 (RGB) 1.96 (RGB ff) 1.93 (PS ff)

\noindent
Upper dimension: 2.06 (PS) 2.01 (RGB) 2.02 (RGB ff) 2.075 (PS ff)

\vspace{2mm}

\noindent
{\em RGB Colour partition:}

\noindent
Black = 000, 001, 011, 101

\noindent
Red = 100

\noindent
White = 010, 110, 111

\vspace{2mm}

\noindent
{\bf Summary:} Satisfies fractal authentication criteria if the more lax force-fitting procedure
is allowed.

\vspace{3mm}

\noindent
{\bf 4. The Wooden Horse: Number 10A, 1948} Pollock (1948) [JPCR 207] (Oil and enamel paint on canvas).

\vspace{2mm}

\noindent
Physical Dimensions: 90.1 $\times$ 190.5 cm

\noindent
Digital Image: 2340 $\times$ 1176 (PS and RGB)

\noindent
Range of Box Sizes: 0.8 cm to 14.2 cm (PS) 0.8 cm to 14.0 cm (RGB) 0.8 cm to 9.5 cm 
(RGB, Orange Layer)

\vspace{2mm}

\noindent
{\em Red Layer:}

\noindent
Area of Layer: 3.6 \% (PS) 2.5 \% (RGB)

\noindent
Noise: 0.023 (PS) 0.026 (RGB)

\noindent
$L_T$: 4.32 cm (PS) 4.32 cm (RGB)

\noindent
Lower dimension: 1.16 (PS) 1.12 (RGB)

\noindent
Upper dimension: 1.56 (PS) 1.63 (RGB)

\vspace{2mm}

\noindent
{\em Orange Layer:}

\noindent
Area of Layer: 1.7 \% (PS) 1.9 \% (RGB)

\noindent
Noise: 0.0301 (PS) 0.027 (RGB)

\noindent
$L_T$: 3.23 cm (PS) 2.67 cm (RGB)

\noindent
Lower dimension: 1.39 (PS) 1.50 (RGB)

\noindent
Upper dimension: 1.18 (PS) 1.25 (RGB)

\vspace{2mm}

\noindent
{\em Yellow Layer:}

\noindent
Area of Layer: 1.2 \% (PS) 1.3 \% (RGB)

\noindent
Noise: 0.026 (PS) 0.018 (RGB)

\noindent
$L_T$: 7.63 cm (PS) 1.67 (RGB)

\noindent
Lower dimension: 1.11 (PS) 1.26 (RGB)

\noindent
Upper dimension: 1.05 (PS) 1.08 (RGB)

\vspace{2mm}

\noindent
{\em Black and Blue Layers:}

\noindent
Area of Layer: 10.3 \% (PS) 2.3 \% (Blue, RGB) 7.4 \% (Black, RGB)

\noindent
Noise: 0.028 (PS) 0.024 (Black RGB) 0.025 (Blue RGB)

\noindent
$L_T$: 3.91 cm (PS) 4.32 cm (Black RGB) 4.32 cm (Blue RGB)

\noindent
Lower dimension: 1.44 (PS) 1.35 (Black RGB) 1.14 (Blue RGB)

\noindent
Upper dimension: 1.92 (PS) 1.79 (Black RGB) 1.73 (Blue RGB)

\vspace{2mm}

\noindent
{\em White Layer:}

\noindent
Area of layer: 5.5 \% (PS) 9.1 \%

\noindent
Noise: 0.031 (PS) 0.026 (PS; trimmed range) 0.026 (RGB)

\noindent
$L_T:$ 3.91 cm (PS) 3.91 cm (RGB)

\noindent
Lower dimension: 1.30 (PS) 1.39 (RGB)

\noindent
Upper dimension: 1.82 (PS) 1.86 (RGB)

\vspace{2mm}

\noindent
{\em RGB Colour Partition:}

\vspace{2mm}

\noindent
Blue = 001, 011,

\noindent
00-00-01, 00-01-01, 01-00-01,

\noindent       
010-010-011, 010-011-010, 010-011-011, 011-010-011, 011-011-011

\noindent       
0100-0100-0101, 0100-0101-0101, 0101-0100-0101

\noindent
black = 010,

\noindent        
00-00-00, 00-01-00, 01-00-00,

\noindent        
010-010-001

\noindent        
0100-0100-0100, 0100-0101-0100, 0101-0100-0100, 0101-0101-0100,

\noindent        
0101-0101-0101

\noindent
Red = 101,

\noindent         
10-00-00, 10-00-01, 10-01-00, 11-00-00, 11-01-00, 11-01-01

\noindent         
111-100-011

\noindent
White = 111

\noindent
Brown = 10-01-01, 10-10-01

\noindent        
011-010-001, 011-011-001

\noindent        
011-010-010, 011-011-010

\noindent        
110-100-010, 110-100-011,

\noindent
Yellow and orange = 10-10-00, 11-10-00, 11-11-00, 11-11-01

\noindent         
110-101-010, 110-101-011, 111-100-010, 111-101-010, 111-101-011

\vspace{2mm}

\noindent
{\em Notes:} In the PS analysis the blue and black layers were analysed as a
composite. In the RGB analysis the horse's head was not fully excluded from the
white layer; thus the PS analysis of the white layer is more accurate. The yellow and
orange layers in the RGB analysis were separated by incorporating spatial information.
If the pixels lay in the appropriate region of RGB and real space they were deemed yellow;
otherwise orange.

\vspace{2mm}

\noindent
{\bf Summary:} Does not satisfy fractal authentication criteria. 
Exhibits anomalous dimensions in the orange and yellow layers.


\vspace{3mm}

\noindent
{\bf 5. Untitled:} Pollock (1950) [JPCR 797] (Enamel on paper).

\vspace{2mm}

\noindent
Physical Dimensions: 28.3 cm $\times$ 150.3 cm

\noindent
Digital Images: 1145 $\times$ 5150 pixels (PS and RGB) 1145 $
\times $ 2908 (RGB - cropped).

\noindent
Range of box sizes: 0.1 cm to 2.7 cm (PS) 
0.1 cm to 4.4 cm (RGB) 0.1 cm to 3.0 cm (RGB - cropped)

\vspace{2mm}

\noindent
{\em Black Layer:}

\noindent
Area of Layer: 5 \% (PS) 5 \% (RGB)

\noindent
Noise: 0.020 (PS) 0.024 (RGB) 0.024 (RGB - cropped)

\noindent
$L_T$: 0.5 cm (PS) 0.5 cm (RGB) 0.5 cm (RGB - cropped)

\noindent
Lower dimension: 1.73 (PS) 1.72 (RGB) 1.72 (RGB - cropped)

\noindent
Upper dimension: 1.36 (PS) 1.37 (RGB) 1.39 (RGB - cropped)

\vspace{2mm}

\noindent
{\em RGB Colour partition:}

\noindent
Black = 000

\vspace{2mm}

\noindent
{\bf Summary:} Does not satisfy fractal authentication criteria. Exhibits anomalous 
dimensions. 

\vspace{3mm}

\noindent
{\bf 6. Number 8, 2007:} Ash and Hallen (2007) (Oil on canvas).

\vspace{2mm}

\noindent
Physical Dimensions: 122 cm $\times$ 152 cm (approx)

\noindent
Digital Image: 1387 $\times$ 1668 pixels (PS) 1192 $\times$ 1566 
pixels (RGB)

\noindent
Range of box sizes: 0.9 cm to 15.3 cm (PS) 0.3 cm to 6.0 cm (RGB)

\vspace{2mm}

\noindent
{\em Black Layer:} 

\noindent
Area of Layer: 44.0 \% (PS) 58.3 \% (RGB)

\noindent
Noise: 0.014 (PS) 0.022 (RGB)

\noindent
$L_T$: 3.8 cm (PS) 1.6 cm (RGB)

\noindent
Lower dimension: 1.89 (PS) 1.81 (RGB)

\noindent
Upper dimension: 2.08 (PS) 2.05 (RGB)

\vspace{2mm}

\noindent
{\em RGB Colour partition:}

\noindent
Black = 000, 001, 010, 011, 100, 101, 110

\noindent
Background = 111

\vspace{2mm}

\noindent
{\em Comments:} The discrepancy between the percentage area of black
obtained by the two colour separation methods is 
is due to the different way the painting was cropped before colour separation.
The cropped image used in the Photoshop analysis 
contained a margin that was largely white. The discrepancy between the
$L_T$ values is presumably due to the same cause.

\vspace{2mm}

\noindent
{\bf Summary:} Satisfies fractal authentication criteria.


\vspace{3mm}

\noindent
{\bf 7. Composition with Red and Black:} Ash and Hallen (2007) (Oil on canvas).

\vspace{2mm}

\noindent
Physical Dimensions: 122 cm $\times$ 152 cm (approx)

\noindent
Digital Image: 1288 $\times$ 1536 pixels

\noindent
Range of box sizes: (PS) 0.3 cm to 16.4 cm (RGB)

\vspace{2mm}

\noindent
{\em Black Layer:}

\noindent
Area of Layer: 33.4 \% (PS) 37.2 \% (RGB)

\noindent
Noise: 0.023 (PS) 0.021 (RGB)

\noindent
$L_T$: 1.75 cm (PS) 1.75 cm (RGB)

\noindent
Lower dimension: 1.68 (PS) 1.64 (RGB)

\noindent
Upper dimension: 2.02 (PS) 2.01 (RGB) 

\vspace{2mm}

\noindent
{\em Red Layer:}

\noindent
Area of Layer: 17.0 \% (PS) 22.3 \% (RGB)

\noindent
Noise: 0.020 (PS) 0.025 (RGB)

\noindent
$L_T$: 4.13 cm (PS) 1.59 cm (RGB)

\noindent
Lower dimension: 1.48 (PS) 1.48 (RGB)

\noindent
Upper dimension: 2.01 (PS) 1.98 (RGB)

\vspace{2mm}

\noindent
{\em RGB Colour partition:}

\noindent
Black = 001, 010, 011,
\newline
00-00-00, 00-00-01, 00-01-00, 00-01-01, 01-01-00, 01-01-01

\noindent
Red = 100, 101, 110,
\newline
      01-00-00, 01-00-01

\noindent
Background = 111

\vspace{2mm}


\noindent
{\bf Summary:} Satisfies fractal authentication criteria.

\begin{figure}[h]
\scalebox{0.8}{\includegraphics{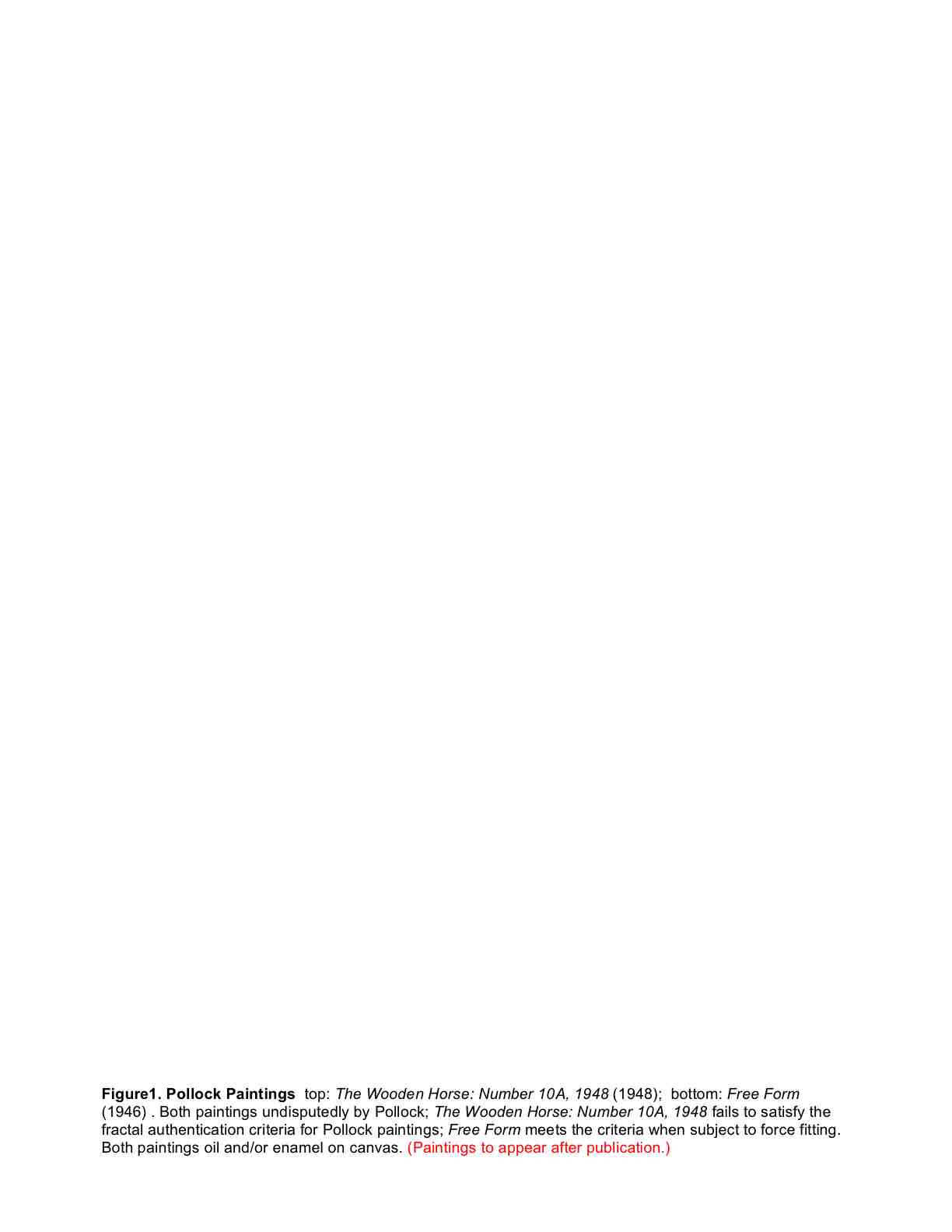}}
\label{gallery2}
\end{figure}

\begin{figure}[h]
\scalebox{0.8}{\includegraphics{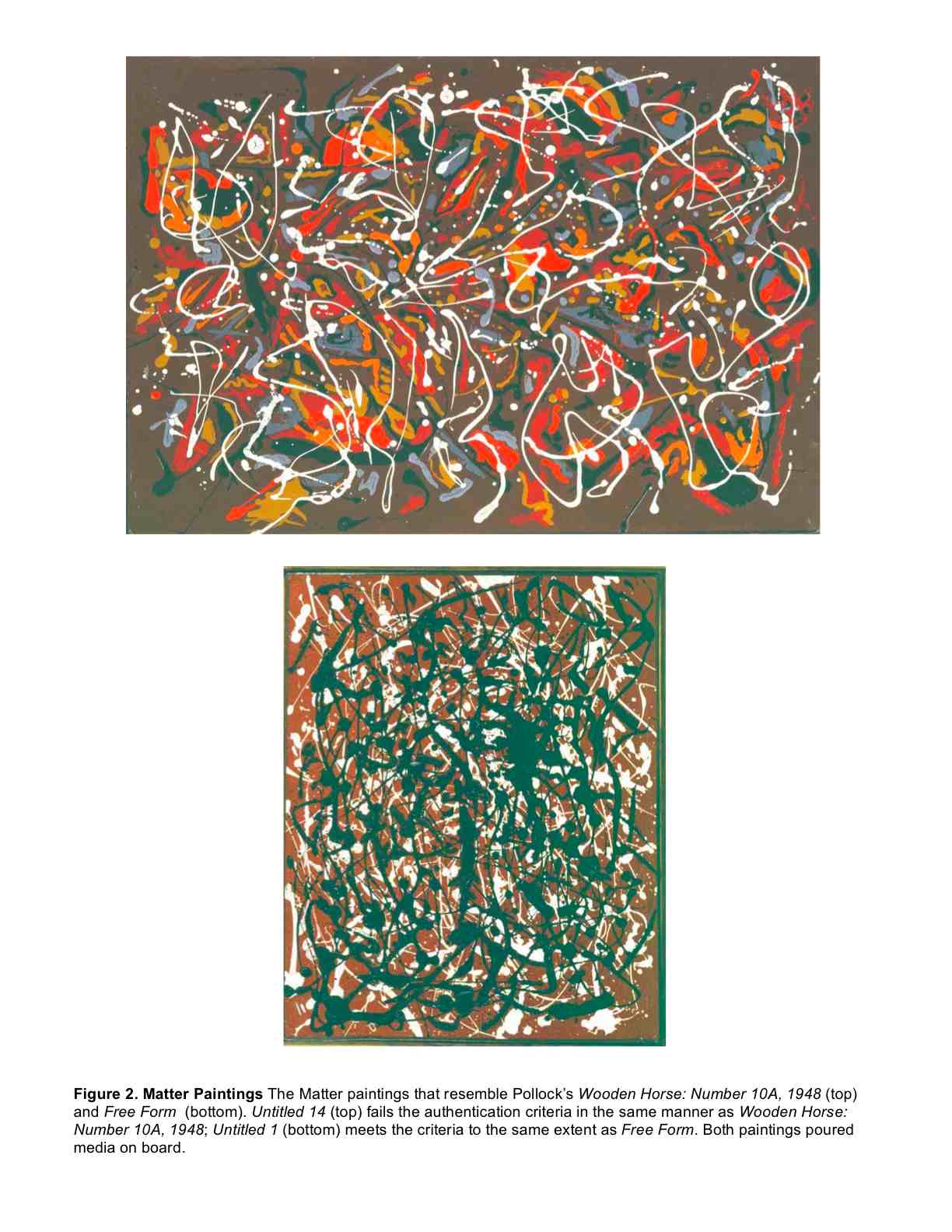}}
\label{gallery2}
\end{figure}

\begin{figure}[h]
\scalebox{0.8}{\includegraphics{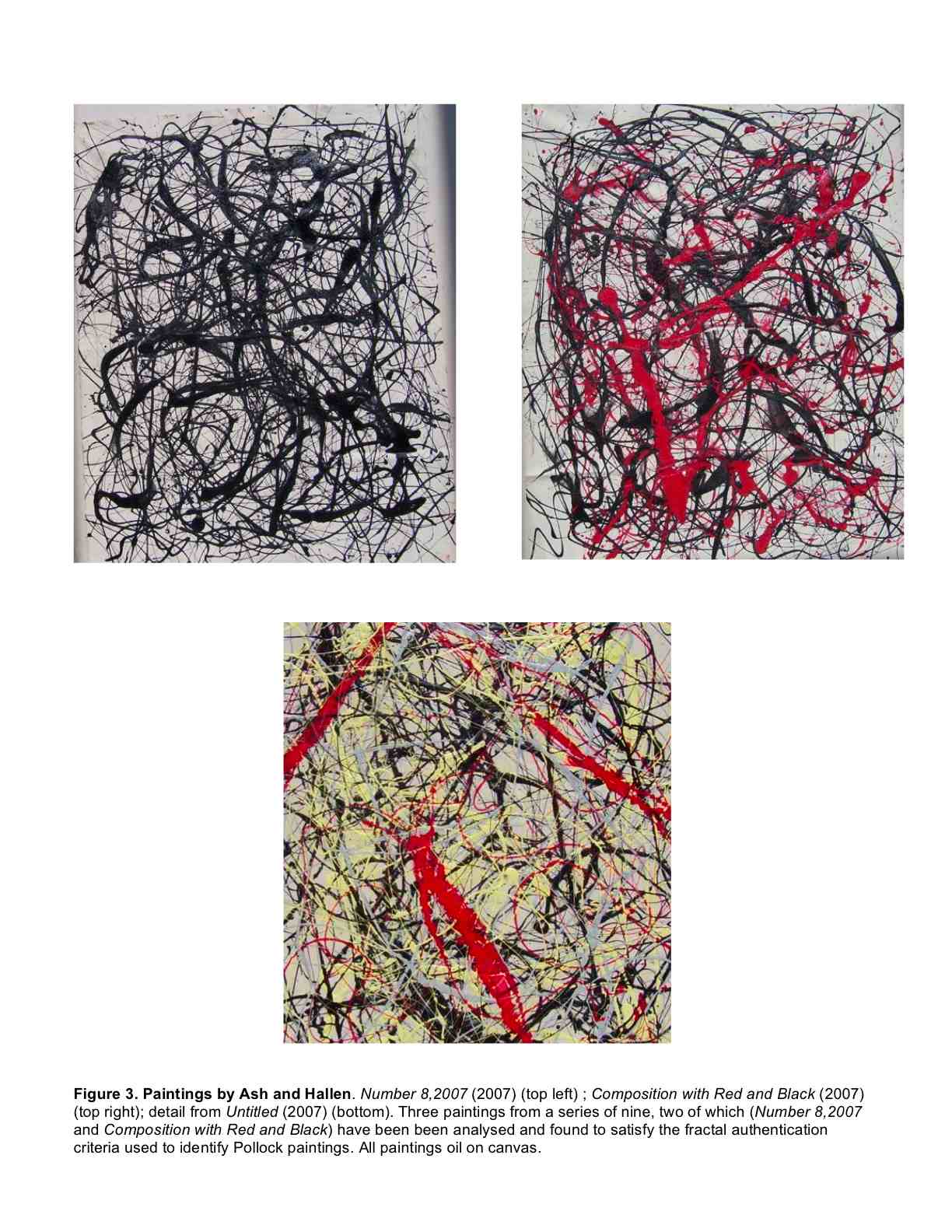}}
\label{gallery2}
\end{figure}

\begin{figure}[h]
\scalebox{0.8}{\includegraphics{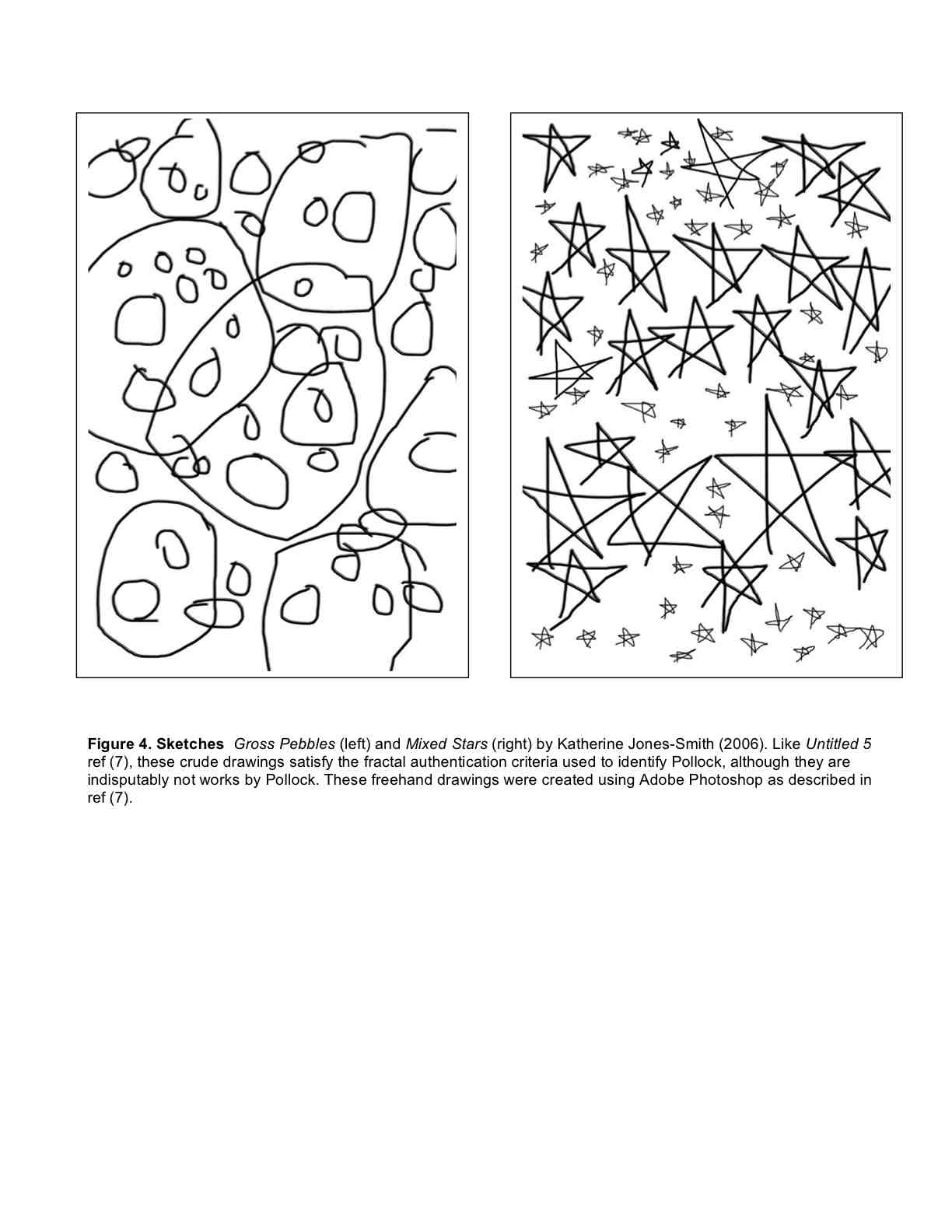}}
\label{gallery2}
\end{figure}

\end{document}